\documentclass[prodmode,acmtompecs]{acmsmall} % Aptara syntax

\usepackage[ruled]{algorithm2e}

\SetAlFnt{\small}
\SetAlCapFnt{\small}
\SetAlCapNameFnt{\small}
\SetAlCapHSkip{0pt}
\IncMargin{-\parindent}
\acmVolume{9}
\acmNumber{4}
\acmArticle{39}
\acmYear{2010}
\acmMonth{3}

\usepackage{amsfonts,amssymb,amsmath}
\usepackage{color}
\usepackage{graphicx}
\usepackage{cite}
\usepackage{url}
\usepackage{bbm}
\usepackage{mathrsfs}
\usepackage{epstopdf}

\usepackage{subfigure}

\DeclareMathAlphabet\mathbfcal{OMS}{cmsy}{b}{n}
\DeclareMathAlphabet{\mathsfbf}{OT1}{cmss}{sbc}{n}

\newtheorem{property}{Property}
\newcommand{\EE}{\mathbb{E}} % average operator
\newcommand{\PP}{\mathbb{P}} % probability
\newcommand{\RR}{\mathbb{R}} % real set
\newcommand{\NN}{\mathbb{N}} % integer set

\newcommand{\ee}{{\rm e}}
\newcommand{\dd}{{\rm\,d}} % differential (for integrals)

\newcommand{\erf}{{\rm erf}} % error function

\newcommand{\av}{{\bf a}}
\newcommand{\bv}{{\bf b}}

\newcommand{\nv}{{\bf n}}

\newcommand{\qv}{{\bf q}}

\newcommand{\vv}{{\bf v}}
\newcommand{\xv}{{\bf x}}
\newcommand{\yv}{{\bf y}}

\newcommand{\zerov}{{\bf 0}}
\newcommand{\onev}{{\bf 1}}

\newcommand{\Am}{{\bf A}}

\newcommand{\Id}{{\bf I}}

\newcommand{\Ym}{{\bf Y}}
\newcommand{\Zm}{{\bf Z}}

\newcommand{\Ac}{{\cal A}}

\newcommand{\Cc}{{\cal C}}

\newcommand{\Nc}{{\cal N}}

\newcommand{\Qc}{{\cal Q}}

\newcommand{\Wc}{{\cal W}}

\newcommand{\Zc}{{\cal Z}}

\newcommand{\betav}{\boldsymbol{\beta}}

\newcommand{\muv}{\boldsymbol{\mu }}

\newcommand{\phiv}{\boldsymbol{\phi}}

\newcommand{\thetav}{\boldsymbol{\theta}}

\newcommand{\Gammam}{\boldsymbol{\Gamma}}

\newcommand{\Sigmam}{\boldsymbol{\Sigma}}

\def\Tran{\mathsf{^T}}

\def\ben{\begin{enumerate}}
\def\beq{\begin{equation}}
\def\beqa{\begin{eqnarray}}
\def\bit{\begin{itemize}}
\def\een{\end{enumerate}}
\def\eeq{\end{equation}}
\def\eeqa{\end{eqnarray}}
\def\eit{\end{itemize}}

\def\non{\nonumber\\}

\begin{document}

\title{Selecting the top-quality item through crowd scoring} 
%\author{}

\author{Alessandro Nordio\affil{CNR-IEIIT, Torino, Italy}
  Alberto Tarable\affil{CNR-IEIIT, Torino, Italy}
  Emilio Leonardi\affil{Politecnico di Torino, Italy; CNR-IEIIT, Torino, Italy}
  Marco Ajmone Marsan\affil{Politecnico di Torino, Italy; CNR-IEIIT, Torino, Italy; IMDEA Networks Institute, Madrid, Spain}
}

\begin{abstract} 
We investigate crowdsourcing algorithms for finding the top-quality item within
a large collection of objects with unknown intrinsic quality values.  This is an
important problem with many relevant applications, for example in networked
recommendation systems.  The core of the algorithms is that objects are
distributed to crowd workers, who return a noisy and biased evaluation.  All
received evaluations are then combined, to identify the top-quality object.  We
first present a simple probabilistic model for the system under investigation.
Then, we devise and study a class of efficient adaptive algorithms to assign in
an effective way objects to workers. We compare the performance of several
algorithms, which correspond to different choices of the design
parameters/metrics. In the simulations we show that some of the algorithms achieve near optimal
performance for a suitable setting of the system parameters.
\end{abstract}

\maketitle

\section{Introduction}

Crowdsourcing is a term often adopted to identify distributed systems that can be
used for the solution of a wide range of complex problems by integrating a large
number of human and/or computer efforts \cite{asurveyofcrowdsourcingsystems}.

The key elements of a crowdsourcing system are: i) the availability of a large
pool of individuals or machines (called {\em workers} in crowdsourcing jargon)
that can offer their (small) contribution to the problem solution by executing a
{\em task}; ii) an algorithm for the partition of the problem at hand into
tasks; iii) an algorithm for the selection of workers and the distribution of
tasks to the selected workers; iv) an algorithm for the combination of workers'
{\em answers} into the final {\em solution} of the problem, v) a {\em requester}
(a.k.a. employer), who uses the three algorithms above to structure his problem
into a set of tasks, assign tasks to selected workers, and combine workers'
answers to obtain the problem solution.

Workers are typically not 100\% reliable, in the sense that they may provide
incorrect answers, and may be biased for different
reasons. Hence, the same task is normally assigned in parallel (replicated) to
several workers, and then a decision rule is applied to their answers. A natural
trade-off between the accuracy of the decision and cost arises; indeed, by
increasing the replication factor of every task, we can increase the accuracy of
the final decision about the task solution, but we necessarily incur higher
costs (or, for a given fixed cost, we obtain a lower task throughput). 

A number of sophisticated software platforms have been recently developed for
the exploitation of the crowdsourcing paradigm.
Some relevant application scenarios
taken from the domains of recommendation and evaluation, are the development of
hotel and restaurant rating systems, the implementation of recommendation
systems for movies, the management of the review process of large conferences.

Given the scale of the current
applications of crowdsourcing systems, the relevance of high-performance and
scalable algorithms is enormous (in some cases, it can have huge economical
impact).

In the examples above, the goal is to find the best, or the $k$ best, elements in a
group of objects in which each object has an intrinsic (unknown) quality
metric. This is a very fundamental algorithmic problem, which has already been
investigated by several researchers in the context of crowdsourcing.

Several papers in the previous literature model workers as only able to
directly compare items in groups comprising two or more objects, expressing a
preference. However, this may not be feasible in many practical scenarios
(e.g. recommendation systems for hotels or restaurants) where users can be
requested to evaluate the last place they have visited.  In this paper, we
focus on the problem of finding the best object within a class and assume that
workers are able to evaluate (in absolute terms) the quality of an object,
providing a noisy score. A similar path was followed in the recent (still
unpublished) work by Khan and Garcia Molina \cite{GMhybrid}, which studies
algorithms to find the maximum element in a group of objects, and discusses
approaches based on comparisons, on ratings, as well as on a mix of the two
possibilities. The main difference between the cited work and this one is in the
quantization of workers' scores.  Indeed, \cite{GMhybrid} assumes that workers'
answers are coarsely quantized over few levels (typically three or five), and
this makes objects with similar quality indistinguishable, so that direct
comparisons and tournaments become necessary to break ties.

On the contrary, we first consider unquantized workers' answers, so as to
maximize the amount of information provided by the workers.  We show that in
this context the scoring approach is superior (in some cases by far) to the
approach based on direct comparisons. This should not be surprising, since
quantization and comparisons entail a partial loss of information.  Then, we show
that by adopting smart quantization techniques with a sufficiently large number
of quantization levels (in the order of few tens) we can closely approach the
performance of systems operating on unquantized scores.

Another significant difference with respect to \cite{GMhybrid} is in the scope
of the works. We aim at the definition of smart multiround {\em adaptive}
algorithms that effectively distribute the resources (workers) among the
objects, at every round, making online decisions whether to distribute further
resources, based on past collected answers. The paper \cite{GMhybrid}, instead,
focuses on non-adaptive algorithms distributing resources to objects according
to a fixed, pre-established scheme.

Our main findings are:
\begin{itemize}
\item resources (workers) must be allocated in a careful manner, concentrating
  more resources on the top-quality objects; this can be done only if algorithms
  are adaptive and, at every round, exploit currently available information
  about objects' quality to decide how to distribute further resources;

\item when workers are affected by bias, accurate bias estimation which can be
  carried out with affordable complexity, can limit performance loss with
  respect to the unbiased case;

\item in several scenarios, algorithms operating on unquantized scores are shown
  to be more efficient than algorithms based on direct comparisons of objects;
  moreover, in such scenarios, tournament-based approaches, that partition
  objects into subgroups and move winners in each sub-group to the next round,
  may become extremely inefficient;
  
\item more practical quantized schemes perform very close to their ideal
  unquantized counterparts, provided that a reasonable number of quantization
  levels is properly assigned to workers' answers.
\end{itemize}

\section{Related work}
\label{sec:RW}
As already said, in most papers dealing with finding the best in a
group of objects, the proposed algorithms consist of comparisons
arranged in rounds, forming a {\em tournament}, and the investigation
concentrates on the trade-offs that appear in this context (e.g.,
cost, accuracy, latency) \cite{GM1,GM2,GM3,GM4,Davidson,siam}.

The problem we study in this paper can be cast in the classical
framework of the exploitation-vs-exploration trade-off~\cite{AzoulaySchwartz20041}. In words,
our adaptive algorithm works in rounds, and, at each round, it has to
choose whether to capitalize on the previous results (favoring those
objects which are currently deemed to be the best candidates for
winning) or to widen its own knowledge (favoring those objects whose
quality is still known with little accuracy). Multiarmed bandit
algorithms~\cite{LAI19854} are a family of algorithms that is typically employed
to solve problems of this type. Briefly, one has several levers to
pull, each of which is characterized by a different reward probability
distribution. Many papers (e.g.,~\cite{10.2307/1427934,Auer2002}) give algorithms for a variety of
scenarios, that tell, in a sequence of pulls, which lever to pull
next, in order to minimize asymptotically the regret, i.e., the
average reward loss with respect to choosing always the (unknown)
optimal lever. A variant for comparisons is the dueling bandit
problem, where one has to pull pairs of levers at a time. In~\cite{Zoghi}, an
algorithm for regret minimization in such a scenario is
described. While bandit problems show similarities with our problem,
there are also some differences. In particular, when we assume that
each worker adds a constant bias to all her answers, to maintain the
hypothesis of independent pulls, we are forced to consider the pair
object-worker as a single lever. However, in our work, we add the constraint that
each pair object-worker is allocated at most once, and, more
importantly, the average reward would be the sum of object quality and
worker bias, which is not what we are interested in. If, on the
contrary, there is no bias, levers can be identified with 
objects, but our problem has still some differences with respect to the
multiarmed bandit problem. First, in our solution, a round corresponds
to the request of a certain number of new evaluations at once, which
is certainly more efficient, if time is taken into account, than
requesting a single evaluation at each round, as it is the case for
bandit problems.  Second and more important, our target is not
precisely to minimize the regret but to reduce the 
probability of incorrectly identifying the best object, which is a more complicate and non-linear function of the
sequence of pulls.

In our paper, workers are distinguished only by bias, while their
evaluation variance (which is directly related to their skills) is
generally assumed to be the same. There are several papers in which
workers in a crowdsourcing environment are assumed to be
indistinguishable (e.g., \cite{NIPS2011_4396,NIPS2012_4701,GM1}). The reason for this lies in the fact that
in many scenarios the workers' skill is difficult to assess, or it is
meaningless, as for example when they are requested to choose the best
picture in a set, to judge in a beauty context, or to provide an
opinion about a restaurant or hotel. In other cases, workers are
selected from prefiltered sets of skilled individuals like the
reviewers of a conference or journal paper. In~\cite{Oh}, the workers are
characterized by different skills (i.e., variance), and a
belief-propagation algorithm is proposed to obtain quality
estimation. However, in~\cite{Oh}, the goal is simply to estimate quality
and not to find the best object, which implies a radical difference in
how the objects are allocated to workers with respect to our problem.
An online allocation strategy for multiclass labeling akin to multiarmed bandit problems is proposed in~\cite{Liu:2015}.

In this paper, in order to find the best of $N$ objects, we opt for
{\em adaptive algorithms} which organize the evaluation in rounds.
In~\cite{NIPS2016} a discussion of the advantage of adaptive task
allocation in crowdsourcing environment is provided, together with
performance guarantees, when workers provide binary answers. A further
paper that compares adaptive and non-adaptive algorithms in different contexts
is~\cite{HoVaughan}. Non-adaptive algorithms for microtask-based crowdsourcing systems
with binary and multilevel answers are proposed in~\cite{NIPS2011_4396,Karger:2013}.

Finally, a problem related to ours is the ranking of $N$ objects. To solve it,
in~\cite{NIPS2012_4701} an algorithm based on pairwise comparisons is
proposed. The workers' answers are used to build a labeled graph on
which a PageRank-like algorithm is employed for discovering the
scores.  

\section{System Assumptions}
\label{sec:SA}
We consider a set of $N$ objects, each of which is endowed with an intrinsic
quality, whose evaluation requires human capabilities.  Let $\xv = [x_1, \dots,
  x_N]$ be the vector of all quality values, which are instances
of the i.i.d. random variables $\qv=[q_1, \dots, q_N]$ having common pdf $f_q$
over $\RR$.

Our goal is to use a crowdsourcing approach to identify the ``best'' object,
that is, the object with the largest quality value, denoted by $x_{i^{*}}$, where
\[ i^* = \arg \max_i x_i\,.\]

To this purpose, a pool of $W$ crowd workers is available. Clearly workers'
evaluations of object quality are prone to errors. We assume, for the moment,
that workers provide absolute unquantized estimates (scores) of the intrinsic
quality of individual objects. We model the error made in such evaluation
process as a (possibly nonzero-mean) additive Gaussian noise.  More precisely,
if the $i$-th object is sent for evaluation to worker $w$, $w = 1,\dots, W$, the
worker's answer will be given by:
\[
a_{iw} = x_i + n_{iw}
\]
where $n_{iw}$ is a Gaussian random variable with mean $b_w$ and variance
$\sigma^2$. Errors on different evaluations are assumed to be independent. Let
$\bv = [b_1, \dots, b_W]$ be the vector of workers' biases, which, as for the
quality values, are supposed to be instances of i.i.d. random variables $\betav
= [\beta_1, \dots, \beta_N]$ having common pdf $f_{\beta}$ over $\RR$.

The error mean reflects the existence of subjective factors influencing quality
assessment for all objects in the same way, whilst the error variance
corresponds to the fact that, in each evaluation, workers may avoid devoting the
effort that is needed for an accurate assessment. Notice that our model assumes
equal error variance for all workers, so that all of them can be considered to
have the same reliability. Although this hypothesis, which we made for the sake
of simplification, is rather strong, our framework still retains all the
conceptual aspects of a more complex worker model.

Observe that our model differs from \cite{GM1,GM2,GM3,GM4,Davidson,siam} because
we assume that workers can provide absolute estimates (scores) of the intrinsic
quality of objects, while \cite{GM1,GM2,GM3,GM4,Davidson,siam} assume workers to
be only able to perform noisy comparisons between groups of objects.  We
wish to remark that the Gaussian model of worker error is in agreement
with Thurstone's law of comparative judgment~\cite{Thurstone}, according to
which comparisons are based on latent quality estimations, which can be well
modeled by Gaussian-distributed random variables. In the context of
crowdsourcing, a similar model has recently been employed also
in~\cite{GMhybrid}.

\begin{figure}[t]
%  \centerline{\resizebox{0.7\columnwidth}{!}{algorithm.ps}}
\centerline{\includegraphics[width=0.7\columnwidth]{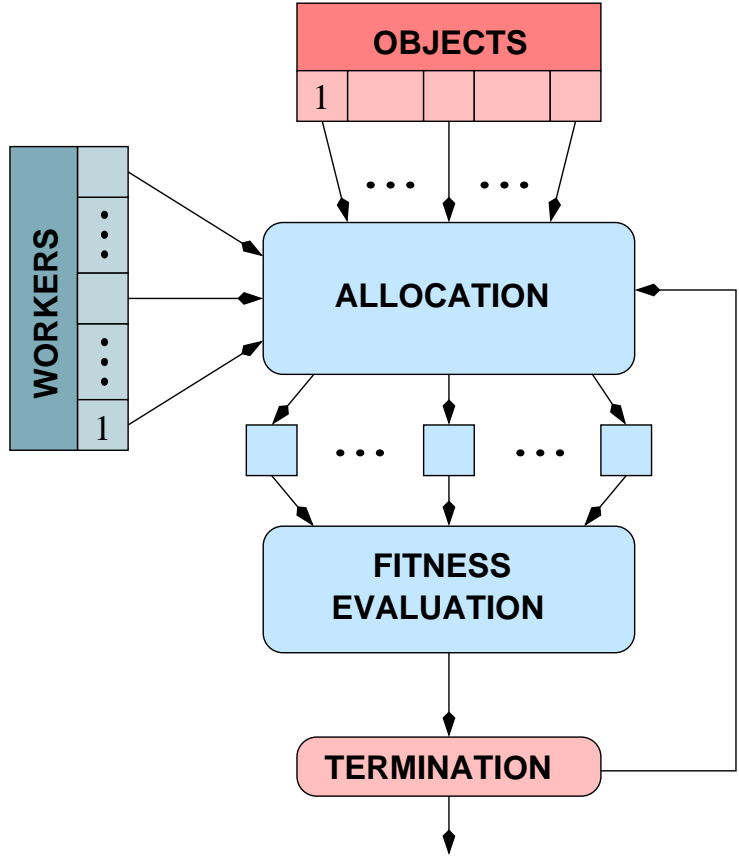}}  
 \caption{Block scheme of the proposed algorithms}
  \label{fig:block_scheme}
\end{figure}

\section{Algorithmic Approach}
\label{sec:AD}
We investigate a class of adaptive algorithms in which objects are sent out for
evaluation through several \emph{rounds}. In each round, each object receives a
given number of evaluations by crowd workers (possibly zero, for some
objects). Then, on the basis of all collected workers' answers, the algorithms
take decisions about the opportunity of requesting extra evaluations for a
subset of the objects in a further round. If no extra evaluations are carried
out, the algorithms terminate and a winner is identified.

More formally, denote with $m^{(\ell)}_i$ the number of evaluations
the $i$-th object has received in round $\ell$ and with $M^{(\ell)}_i$
the total number of evaluations received by object $i$ up to (and
including) round $\ell$. We define $\av_i^{(\ell)} =
\left[a_{ij}^{(\ell)}\right]$, $j=1,\ldots,M_i^{(\ell)}$ as the
vector\footnote{In order to avoid cumbersome notation we sometimes
  indicate the vector $\vv=[v_1,\ldots,v_n]$, as $\vv=[v_i]$,
  $i=1,\ldots,n$} of random variables representing the answers about
object $i$ collected up to round $\ell$, and $\Am^{(\ell)} =
\left[\av_1^{(\ell)}, \dots, \av_N^{(\ell)}\right]$.

At the beginning of round $\ell$, with $\ell\ge 1$, the fitness index
$\phi_i^{(\ell-1)}$ represents a metric associated with the quality of
the object $i$,  as a result of processing previous evaluations.  If
$\phi_i^{(\ell-1)}>\phi_j^{(\ell-1)}$ then the quality of object $i$
is estimated to be larger than the quality of object $j$.  Let the
vector of all fitness indices be
$\phiv^{(\ell-1)}= [\phi_1^{(\ell-1)}, \ldots,\phi_N^{(\ell-1)}]$.
For $\ell=1$, i.e., when no evaluations are available yet, fitness
indices are equal for all objects, since the object qualities are
assumed to be instances of i.i.d random variables.

In round $\ell$, some of the objects may have a fitness index equal to
$-\infty$. These objects are not assigned any further evaluation and are out of
the contest. Define the contestant set $\Cc^{(\ell)}$ at round $\ell$ as the set of
objects for which the fitness index is currently larger than $-\infty$, i.e.,
\[
\Cc^{(\ell)} = \left\{i \in \{1,\dots,N\}:  \phi_i^{(\ell-1)} > -\infty\right\}\,.
\]
We remark that, according to our algorithms, the contestant set at
round $\ell+1$ is always a (possibly improper) subset of the
contestant set at round $\ell$, i.e., $\Cc^{(\ell+1)}\subseteq
\Cc^{(\ell)}$.

On the basis of $\phiv^{(\ell-1)}$ and, possibly, of the total number
of past assignments ${M}^{(\ell-1)} = \sum_{i=1}^N M_i^{(\ell-1)}$,
the algorithm decides, according to a termination rule, whether to
stop or to go on with the rounds. If rounds are stopped, the object
with the largest fitness index is declared the winner.  Otherwise, a
budget of new worker evaluations is assigned to objects. Such budget
is dimensioned as
\[
[m_1^{(\ell)}, \dots, m_N^{(\ell)}] = \Ac(\phiv^{(\ell-1)}, M^{(\ell-1)})
\]
where $\Ac(\cdot)$ is the $\RR^{N} \times \NN \to \NN^N$ allocation function. We assume
function $\Ac(\cdot)$ to be increasing with respect to the object quality, i.e.,
our algorithm tends to allocate more workers to objects with top estimated
quality, as formally stated below.
\begin{property}
If $ \phi^{(\ell-1)}_i < \phi^{(\ell-1)}_j$, then $m_i^{(\ell)} \le m_j^{(\ell)}$,
with $m_i^{(\ell)}=0$ if $\phi^{(\ell)}_i=-\infty$.
\end{property}

After determining the number of suitable evaluations, the objects are sent to
available crowd workers, according to a suitable worker selection policy, and
answers are collected.  Then, such answers, together with the previous ones, are
used to update the fitness vector for next round, $\phiv^{(\ell)}$ (and,
consequently, $\Cc^{(\ell+1)}$).  Several algorithms can be devised in
accordance to the previous scheme, depending on how we select the different
metrics and parameters, such as $\Ac(\cdot)$, the fitness index, the worker
selection policy and the termination rule.

A graphical representation of the proposed algorithms is depicted in
Figure~\ref{fig:block_scheme}.

\section{Design Parameters}
\label{sec:DP}
In this section we design the parameters of the algorithms under the assumption
that workers' answers are unquantized.

\subsection{A-posteriori quality estimation}
At round $\ell$ of the algorithm, on the basis of the collected answers we can
compute an a-posteriori distribution $f_{\qv}^{(\ell)}(\xv | \Ym^{(\ell)})$ for
the quality vector $\qv$, where $\Ym^{(\ell)}$ represents a realization of the
random vector $\Am^{(\ell)}$ and collects all the answers up to round $\ell$.

In order to find $f_{\qv}^{(\ell)}(\xv | \Ym^{(\ell)})$, let us rewrite in
matrix form the relationship between the quality values, the biases and the
workers' answers as
\begin{equation}
\Ym^{(\ell)} = \left[ 
\begin{array}{cc}
\Gammam^{(\ell)}_{\xv} & \Gammam^{(\ell)}_{\bv} 
\end{array}\right] \left[
\begin{array}{c}
\xv \\
\bv
\end{array} \right]
+ \nv^{(\ell)}
\end{equation}
where 
\begin{itemize}
\item $\Gammam^{(\ell)}_{\xv}$ is a $M^{(\ell)} \times N$ binary allocation matrix, whose $j$-th row has a single 1, located in column $i$ if the $j$-th answer in $\Ym^{(\ell)}$ is an evaluation for object $i$;

\item $\Gammam^{(\ell)}_{\bv}$ is a $M^{(\ell)} \times W$ binary allocation matrix, whose $j$-th row has a single 1, located in column $w$ if the $j$-th answer in $\Ym^{(\ell)}$ is an evaluation of worker $w$; 

\item $\nv^{(\ell)}$ is a vector of uncorrelated Gaussian random variables with zero mean and variance $\sigma^2$.
\end{itemize}

Let $\Gammam^{(\ell)} = \left[ 
\begin{array}{cc}
\Gammam^{(\ell)}_{\xv} & \Gammam^{(\ell)}_{\bv} 
\end{array}\right]$ be the overall allocation matrix and $\thetav = \left[
\begin{array}{cc}
\xv\Tran & \bv\Tran
\end{array} \right]\Tran$ be the set of all parameters to be estimated. Thanks to Bayes' rule, the joint a-posteriori 
distribution of $\qv$ and $\betav$ can be written as:
\beq
f^{(\ell)}_{\qv,\betav}(\thetav | \Ym^{(\ell)})
\propto 
f({\Ym}^{(\ell)} | \thetav) \prod_{i=1}^{N} f_q(x_i) \prod_{w=1}^{W} f_{\beta}(b_w)\,.
\eeq
Let $\Ym^{(\ell)} = \left[ \widetilde{\Ym}^{(1)}\Tran, \dots, \widetilde{\Ym}^{(\ell)}\Tran \right]\Tran$ where $\widetilde{\Ym}^{(i)}$
are the answers collected at round $i$ only. Similarly let $\Gammam^{(\ell)} = \left[ \widetilde{\Gammam}^{(1)}\Tran, \dots, \widetilde{\Gammam}^{(\ell)}\Tran \right]\Tran$ where $\widetilde{\Gammam}^{(i)}$ is the allocation matrix for round $i$ only.
Then by using the chain rule we can write
  \begin{eqnarray*}
    f(\Ym^{(\ell)} | \thetav)
 & = & f(\widetilde{\Ym}^{(1)}, \dots, \widetilde{\Ym}^{(\ell)} | \thetav)\\   
 & = & \prod_{i=1}^{\ell} f(\widetilde{\Ym}^{(i)} | \thetav, \Ym^{(i-1)})
  \end{eqnarray*}
Since $\widetilde{\Ym}^{(i)}$ depends on  $\Ym^{(i-1)}$ through $\widetilde{\Gammam}^{(i)}$ we can replace  $\Ym^{(i-1)}$ by $\widetilde{\Gammam}^{(i)}$ in the previous equation. It follows that
  \begin{eqnarray*}    
    f(\Ym^{(\ell)} | \thetav)
    & = & \prod_{i=1}^{\ell} f(\widetilde{\Ym}^{(i)} | \thetav, \widetilde{\Gammam}^{(i)}) \\
& \propto & \prod_{i=1}^{\ell} \exp \left\{-\frac{\|\widetilde{\Ym}^{(i)} - \widetilde{\Gammam}^{(i)}\thetav \|^2}{2\sigma^2}\right\} \\
%& = & \exp \left\{-\frac{\sum_{i=1}^{\ell} \|\widetilde{\Ym}^{(i)} - \widetilde{\Gammam}^{(i)}\thetav \|^2}{2\sigma^2}\right\} \\
& = & \exp \left\{-\frac{ \|\Ym^{(\ell)} - \Gammam^{(\ell)}\thetav \|^2}{2\sigma^2}\right\} 
  \end{eqnarray*}
Therefore
\begin{equation}
f^{(\ell)}_{\qv,\betav}(\thetav | \Ym^{(\ell)})
= \kappa 
  \ee^{-\|\Ym^{(\ell)} \mathord{-} \Gammam^{(\ell)} \thetav \|^2/(2\sigma^2)} \prod_{i=1}^{N} f_q(x_i) \prod_{w=1}^{W} f_{\beta}(b_w)
\nonumber
\end{equation}
where $\kappa$ is such that $\int f^{(\ell)}_{\qv,\betav}(\thetav |
\Ym^{(\ell)}) \dd \thetav=1$.

When the a-priori pdf's $f_q$ and $f_{\beta}$ are
Gaussian with means $\mu_{q}$ and $\mu_{\beta}$, and variances $\sigma_{q}^2$
and $\sigma_{b}^2$, respectively, as a consequence of the fact that the Gaussian
distribution is self-conjugate with respect to the Gaussian likelihood function,
we easily obtain that $f^{(\ell)}_{\qv,\betav}$ is also Gaussian, with
covariance matrix
\begin{equation}\label{eq:cov_mat}
\Sigmam_{\qv,\betav}^{(\ell)} = \left( \frac{\Gammam^{(\ell)}\Tran \Gammam^{(\ell)}}{\sigma^2} + \left(\Sigmam_{\qv,\betav}^{(0)}\right)^{-1} \right)^{-1}
\end{equation}
and mean
\begin{equation}\label{eq:mean_vec}
\muv_{\qv,\betav}^{(\ell)} = \Sigmam_{\qv,\betav}^{(\ell)}  \left( \frac{\Gammam^{(\ell)}\Tran \Ym^{(\ell)}}{\sigma^2} + \left(\Sigmam_{\qv,\betav}^{(0)}\right)^{-1} \muv_{\qv,\betav}^{(0)}\right)
\end{equation}
where 
\[
\Sigmam_{\qv,\betav}^{(0)} = \left[
\begin{array}{cc}
\sigma^{2}_{q} \Id_{N} & \zerov \\
\zerov & \sigma^{2}_{\beta} \Id_{W} 
\end{array}
\right]\,,\hspace{4ex} 
\muv_{\qv,\betav}^{(0)} = \left[
\begin{array}{c}
\mu_{q} \onev_{N}  \\
 \mu_{\beta} \onev_{W} 
\end{array}
\right]\,, 
\]
being $\Id_K $ the size-$K$ identity matrix and $\onev_{K}$ the length-$K$
all-one column vector. It is worth noting that $\muv_{\qv,\betav}^{(\ell)}$ is
the Minimum-Mean-Square-Error estimate of $\thetav$ given that $\Ym^{(\ell)}$ is
the realization of $\Am^{(\ell)}$.  Observe
that~\eqref{eq:cov_mat}-\eqref{eq:mean_vec} also hold for unknown prior
distribution, in which case we set $\sigma_{x}=\sigma_{b}=\infty$.  Moreover, we
remark that~\eqref{eq:cov_mat}-\eqref{eq:mean_vec} can also be employed to
obtain an approximate a-posteriori distribution in the case of non-Gaussian
priors with given means and variances.

From the above joint a-posteriori distribution of $\qv$ and $\betav$, it is straightforward to obtain the marginal a-posteriori distribution of
$\qv$ in round $\ell$, i.e., $f_{\qv}^{(\ell)}(\xv | \Ym^{(\ell)})$. For
Gaussian priors, $f_{\qv}^{(\ell)}(\xv | \Ym^{(\ell)})$ is Gaussian, with mean
$\muv_{\qv}^{(\ell)} = \left[ \mu_{q_1},\dots,\mu_{q_N}\right]$ and covariance
matrix $\Sigmam_{\qv}^{(\ell)}$ that are equal to the upper part of
$\muv_{\qv,\betav}^{(\ell)}$ and to the upper-left corner of $
\Sigmam_{\qv,\betav}^{(\ell)}$, respectively.

\subsection{Possible performance parameters} 
\label{sec:decision}
In order to properly choose the metrics of the crowdsourcing algorithm, it is
important to identify the performance parameters we may want to
optimize. Several options can be devised. In the following, we will omit the
round index $\ell$ for ease of notation.  Consider an answer realization $\Ym =
[\yv_1,\dots,\yv_N]$ and define the corresponding estimate $\widehat{i^*}(\Ym)$
of $i^*$, in the following simply denoted by $\widehat{i^*}$. A set of possible
performance parameters that can be considered are:
\begin{itemize}
\item the \emph{order-$k$ distortion} $D^{(k)} = \EE
  \left|q_{\widehat{i^*}}-q_{i^*}\right|^k$, $k\in \NN$, which is averaged with
  respect to the current a-posteriori distribution of $\qv$ given $\Am = \Ym$.
  Unfortunately the computation of the distortion is in general too complex,
  even for moderate values of $N$.
\item the \emph{error probability} $p_{\rm e} = \PP \left\{\widehat{i^*}
  \neq i^* \right\}$.  With such a choice, a maximum-a-posteriori (MAP) rule
turns out to be optimal. Precisely, let $\pi_i = \PP\{i^* = i | \Ym\}$
be the probability of $i$ to be the top-quality object, given the
  answers $\Ym$, for $i \in \Cc$, which is the contestant set. We can in
  principle compute the value of $\pi_i$  as
\begin{eqnarray}\label{eq:max_distribution}
\pi_i
&=& \PP\left\{\cap_{j \in \Cc \setminus \{i\}} \{ q_j \mathord{<} q_i\} | \Ym\right\} \non
&=&\int_{-\infty}^{+\infty} \left( \int_{-\infty}^{x_i} \cdots \int_{-\infty}^{x_i} f_{\qv}(\xv | \Ym) \mathbf{\dd}^{N-1} \xv_{\sim i}\right) \dd x_i 
\end{eqnarray}
where $\xv_{\sim i}$ is obtained by removing the $i$-th component from $\xv$.
It is easy to see that $p_{\rm e}$ is minimized when $\widehat{i^*} = \arg \max_i \pi_i$.

Since the evaluation of $\pi_i$, for $i=1,\ldots, N$, entails a
computational complexity growing linearly with $N$, we propose the following
approximation:
\begin{eqnarray}\label{approx-fit}
\widetilde{\pi}_i &=&\int_{-\infty}^{+\infty} \left( \int_{-\infty}^{x} f_{q_i, q_{c(i)}}(x,x' | \Ym) \dd x' \right) \dd x 
\end{eqnarray} 
where $c(i) = \arg \max_{j \neq i} \mu_{q_j}$ corresponds to the object with
maximum current estimated quality except $i$.  In practice, \eqref{approx-fit}
restricts the comparison to only two objects, the running candidate $i$ and its
strongest competitor $c(i)$, and uses the current probability
$\widetilde{\pi}_i(\Ym)$ that object $i$ is better than $c(i)$ as an
approximation for $\pi_i(\Ym)$.  Therefore, by construction, $\widetilde{\pi}_i \geq {\pi}_i$, and the difference decreases as the number of objects with a good estimated quality decreases, as in the last rounds of the algorithm. In the case of Gaussian or unknown
  priors, \eqref{approx-fit} becomes
\begin{equation}
\widetilde{\pi}_i = \frac{1}{2} \left[ 1 + \erf\left(\frac{  \mu_{q_i} - \mu_{q_{c(i)}}}{\sqrt{2 \left( \sigma_{q_i}^{2}+\sigma_{q_{c(i)}}^{2} - 2 \rho_{i,c(i)} \sigma_{q_i} \sigma_{q_{c(i)}} \right)} }\right)\right]
\end{equation}
where $\sigma_{q_i}^2$ and $\rho_{i,j}$ are the current a-posteriori variance
for $q_i$ and correlation coefficient between $q_i$ and $q_j$, respectively.
\end{itemize}
In this work, because of complexity considerations, we choose the {\em error
  probability} as the performance parameter. Therefore our algorithms will be
designed in order to reduce the error probability as much as possible.

\subsection{Fitness indices}
Different choices for fitness indices are possible.

\begin{itemize}
\item {\bf{Exact max probability:}} we identify the fitness index of objects
  with their estimated probability of being the top-quality object:
  $\phi_i=\pi_i$ whose expression is given in~\eqref{eq:max_distribution}.
\item {\bf{Approximate max probability:}} $\phi_i=\widetilde{\pi}_i$ whose
  expression is given in~\eqref{approx-fit}.
\item {\bf{Exact max probability with elimination:}} As stated in the previous
  section, the contestant set, $\Cc$, initially set to $\{1,\dots,N\}$, may be
  shortened along rounds. We have considered a strategy where, at each round,
  those objects whose $\pi_i$ is lower than a threshold $\pi_{\rm th,E}$ are
  eliminated. For this strategy, the fitness index is given by:
\begin{equation}
\phi_i= \left\{ 
\begin{array}{rl}
\pi_i, &  \pi_i > \pi_{\rm th,E} \text{ and } i \in \Cc  \\
-\infty, & \pi_i \leq \pi_{\rm th,E}  \text{ or } i \notin \Cc  
\end{array} \right.
\label{eq:fitness_exact} 
\end{equation}

\item {\bf{Approximate max probability with elimination:}} we can
  consider a strategy where objects are eliminated if
  $\widetilde{\pi}_i$ falls below a threshold $\pi_{\rm th,E}$. The
  corresponding fitness index is given by~\eqref{eq:fitness_exact} where 
$\pi_i$ is replaced by $\widetilde{\pi}_i$.
\end{itemize} 

All these fitness indices have been tested numerically.

\subsection{Allocation function}
As stated in previous sections, given the current fitness index, the allocation
function $\Ac(\cdot)$ determines the number of further evaluations needed by
each object in round $\ell$. Furthermore, $\Ac(\cdot)$ is a non-decreasing function of
the fitness index (Property 1).

For simplicity, we are particularly interested in the case where $\Ac(\cdot)$ returns
values in $\{0,1\}^N $, i.e., where the number of workers assigned to every
object within a round is either 0 or 1. In such a case, in round $\ell$, the
$B^{(\ell)}\le N$ top-quality objects will receive an extra worker, while all
other objects will not receive any extra worker.

Two possible choices are considered in this paper, according to whether the total
evaluation budget is either fixed or not.

\begin{itemize}
\item {\bf{Unbounded budget:}} If there is no maximum number of requested
  evaluations, $\Ac$ only depends on the fitness index, in the following way:
$m_i^{(\ell)}=1$ if  $\phi_i^{(\ell)}  > \pi_{\rm th,A}$ and $m_i^{(\ell)}=0$ otherwise.
Here $\pi_{\rm th,A}$ is a suitable accuracy threshold, generally different
from $\pi_{\rm th,E}$ defined in the previous subsection. However, for
consistency, we need to have $\pi_{\rm th,E} \leq \pi_{\rm th,A}$.  
It is worth noting that the value of $\pi_{\rm th,A}$ determines the trade-off 
between exploitation and exploration: a large value of $\pi_{\rm th,A}$ 
leads to concentrate the evaluations on objects with a high fitness index (exploitation), 
while a small $\pi_{\rm th,A}$ will spread equally the evaluations over most objects (exploration). 
A similar comment applies to the choice of $\pi_{\rm th,E}$.

\item {\bf{Bounded budget:}} If at most $M_{\rm max}$ evaluations can be
  requested in all rounds, then, in round $\ell$, $\Ac(\cdot)$ must take into
  account also the number of evaluations already requested, given by
  $M^{(\ell-1)}$. Let again $\pi_{\rm th,A}$ be the threshold against which the
  fitness index is compared, like in the unbounded-budget case, and let
  $B^{(\ell)}$ be the number of objects that currently pass the threshold. If
  $B^{(\ell)} \leq M_{\rm max}-M^{(\ell-1)}$, then the allocation of new
  evaluations is the same as for unbounded budget, otherwise only the $M_{\rm
    max}-M^{(\ell-1)}$ objects with the largest fitness index are allocated a
  further evaluation.
\end{itemize}

\subsection{Worker selection}
Observe that when either worker's bias is negligible, or it is perfectly
known by the system, workers are indistinguishable. Any possible non-degenerate
(i.e. according to which every object is evaluated at most once by every worker)
allocation of objects to workers will lead to the same performance.  When 
bias comes to play a significant role, things become more involved.  Restricting
our discussion to the case in which workers' bias is completely unknown to
the system before the allocation, we can make the following considerations:

\begin{itemize}
 \item to improve the accuracy of the workers' bias estimate, the worker
   selection strategy should maximize the number of evaluation performed by
   every involved worker, while minimizing the number of involved workers. In
   this way, indeed, on the one hand we minimize the number of parameters to be
   estimated, while, on the other hand, we maximize the number of available
   samples for every parameter to be estimated.
 
 \item worker bias plays an insignificant role in the case in which, at every
   round $l$, all the contestant objects in $\Cc^{(l)}$ have been evaluated by
   exactly the same set of workers and answers are unquantized.  This because,
   in this case, for the cumulative effect of the bias of workers, all object
   evaluations are subject to the same shifts and, therefore, the relative merit
   among objects (i.e differences among object qualities) are not impacted by
   worker bias.
 
 The worker selection strategy we propose is inspired by previous
 considerations.  We assume that every worker can evaluate a limited number of
 objects. Let us denote with $O_{\rm max}$ the maximum number of allocations for
 every worker (for the sake of simplicity we assume $O_{\rm max}$ to be the same
 for every worker, but the extension to the more general case is pretty
 straightforward).  Furthermore, to allow workers to better plan their work, we
 require that every worker receives all the allocations in a unique batch (i.e.,
 the same worker cannot be employed over multiple rounds).  Our allocation
 policy works as follows: as long as $\Cc^{(l)}\le O_{\rm max}$, at every round
 we allocate all the contestant objects to a randomly selected new worker.  As
 long as $\Cc^{(l)}> O_{\rm max}$, instead, we select $W^{(l)}=\lceil
 \Cc^{(l)}/O_{\rm max}\rceil$ new workers and we allocated to every worker
 either $\lfloor \Cc^{(l)}/W^{(l)}\rfloor$ or $\lceil \Cc^{(l)}/W^{(l)}\rceil$
 randomly selected objects so to allocate all the contestant objects to the
 workers.
\end{itemize}

\subsection{Termination rules}
Based on the choice of the fitness index and the allocation function $\Ac$, the
algorithm termination rule may be different.

\begin{itemize}
\item {\bf{Maximum budget achieved:}} When a maximum of $M_{\rm max}$
  evaluations is allowed, reaching this maximum budget will cause rounds to
  stop.

\item {\bf{Singleton contestant set:}} For algorithms that eliminate objects
  when their fitness is lower than $\pi_{\rm th,E}$, the natural termination
  condition is when the contestant set only contains a single object, i.e.,
  $|\Cc^{(\ell)}| = 1$.

\item {\bf{Accuracy:}} If only a single object passes the accuracy threshold
  $\pi_{\rm th,A}$, while all other objects do not, meaning that there is
  already a strong candidate winner, the algorithm may terminate rounds.
\end{itemize}  

When applicable, the termination rule can be the combination of all three rules
above, i.e., the algorithm may terminate whenever one of the three becomes true.

\section{Scoring Versus Direct Comparisons}\label{sec:sco-vs-T}
The goal of this section is to provide a critical comparison between the
performance of algorithms based on (noisy) quantitative estimates of the object
quality and algorithms just resorting to direct comparisons among subsets of
objects. In order to do this, we focus in this section on a toy scenario for which the analysis is tractable. However, we remark that such scenario includes the most significant features of a more complex case.
 
  We start focusing on a simplified scenario in which worker bias can be
  negligible and we show that, in such a case, the former class of algorithms
  appears clearly more effective than the latter.  Let us first consider a toy
  case in which only two objects are given, with qualities $x_1$ and
  $x_2=x_1+\Delta$, and compare two algorithms requiring the same amount of
  human effort. The first algorithm resorts to outcomes of direct comparisons
  between the objects performed by the crowd workers, while the second exploits
  quantitative estimates of the object qualities provided by the same (or other)
  crowd workers.

Observe that an algorithm exploiting the outcomes of direct comparisons between
objects, and employing a fixed budget of $W$ workers for each comparison,
necessarily works as follows. Each of the $W$ enrolled workers returns a binary
variable, indicating which object she prefers. Once all answers, collectively
denoted $\Zm$, are obtained, a majority rule is applied by the algorithm to
choose the ``best'' object. According to our model, each worker prefers object
$1$, if she estimates that the quality of object $1$ exceeds the quality of
object $2$ and vice versa.  Thus, a worker chooses object 1, i.e., returns an
incorrect answer, with probability $p_{\Delta}= \frac{1}{2}
\text{erfc}(\frac{\Delta}{2\sigma})$, while she chooses object $2$, thus
returning a correct answer, with probability $1-p_{\Delta}$.

Processing the $W$ collected answers (to simplify the analysis, we assume an odd
value for $W$), the algorithm based on comparisons erroneously selects object
$1$ whenever the number of answers equal to 1 exceeds $W/2$, i.e.:
\[ p^{\text{comp}}_{\rm e} = \PP\left(\text{Bin}(W, p_{\Delta})>W/2\right) \]
where $\text{Bin}(W,p)$ denotes a binomial distribution of parameters $W$ and
$p$.

\begin{figure}[ht]
\centerline{\includegraphics[width=0.7\columnwidth]{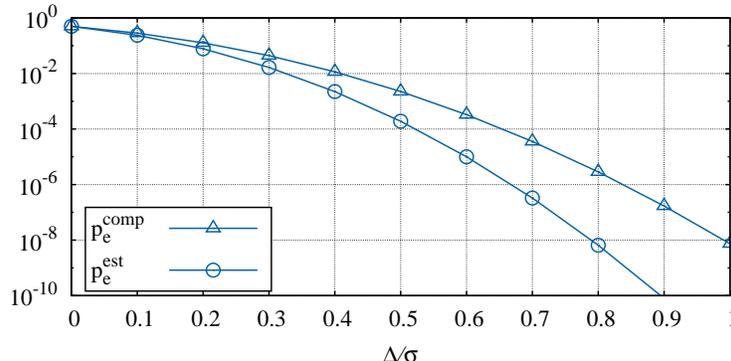}}
\vspace{-2ex}
\caption{Comparison between quantitative
  quality estimations and direct comparisons in terms of error probability in
  the case of two objects. $W=101$.}\label{fig:figure_comp}
\end{figure}

Instead, an algorithm that has access to the quantitative quality estimates
$\Ym$ provided by the $W$ crowd workers, naturally selects the object with the
largest estimated quality $\widehat{x}(\yv_i)$. In this case the error
probability is given by:
\[ p^{\text{est}}_{\rm e}= \frac{1}{2} \text{erfc} \left(\frac{\sqrt{W}\Delta}{2\sigma} \right)\,. \]
Figure~\ref{fig:figure_comp} shows that $p^{\text{est}}_{\rm e} <
p^{\text{comp}}_{\rm e}$. Indeed, from an information-theoretic perspective, the
answers $\Ym$ provide much more information on the quality of the two objects
than the comparisons $\Zm$. This implies that any algorithm resorting to direct
comparisons does not fully exploit the information on object quality that crowd
workers are able to provide and, as a consequence, turns out to be suboptimal.

We now analytically compare $p^{\text{comp}}_{\rm e}$ and $p^{\text{est}}_{\rm
  e}$ to better quantify the advantages of the approach exploiting quantitative
estimates of object quality.  To this end, we approximate the binomial
distribution $\text{Bin}(W, p_{\Delta})$ with a Gaussian distribution
$\Nc(Wp_{\Delta}, \sqrt{W p_{\Delta}(1-p_{\Delta}}))$. By the
De~Moivre-Laplace theorem, such approximation is asymptotically tight for large
$W$.  Following this approach, $p^{\text{comp}}_{\rm e}$ can be approximated as:
\[
p^{\text{comp}}_{\rm e} \approx \frac{1}{2}\text{erfc}\left(\frac{\sqrt{W}(1- 2p_{\Delta})}{2\sqrt{2p_{\Delta}(1-p_{\Delta})}}\right)
\]
To further simplify the expression of $p^{\text{comp}}_{\rm e}$, we consider the
limit for $\frac{\Delta}{\sigma}\to 0$, in which case $p_{\Delta}=\frac{1}{2}\left(1-
\frac{2}{\sqrt{\pi}}\frac{\Delta}{2\sigma}\right)+o(\frac{\Delta}{\sigma})$, and thus
\[
p^{\text{comp}}_{\rm e} \approx \frac{1}{2}\text{erfc}\left(\sqrt{\frac{2}{\pi}}\frac{\sqrt{W}\Delta}{2\sigma}+ o\left(\frac{\Delta}{\sigma}\right) \right)
\]
The performance penalty entailed by the approach resorting to direct comparisons
of objects is expressed by the factor $\sqrt{2/\pi}$ appearing in the argument
of the above erfc function.  This factor has a strong impact on the algorithm
performance, since, for $\Delta/\sigma$ sufficiently small, as $W$ increases,
the ratio $p^{\text{est}}_{\rm e}/ p^{\text{comp}}_{\rm e}$ tends exponentially
fast to zero.

Now, consider a case in which $N > 2$ objects are to be evaluated. For example,
let us focus on the case $N=4$. To declare a winner through direct comparisons
of objects, we need to compare at least three pairs of objects. A natural
solution is to arrange a~\emph{tournament} in which the four objects are first
partitioned into two pairs, so that the objects in each pair can be compared in
parallel (first round). The two winners of the first round are then compared to
identify the globally best object (second round).  Observe that the outcomes
$\Zm$ returned by crowd workers within the first round cannot be exploited in
any way at the second round. In other words, at the end of the first round, no
useful information is available to rank the two first-round winners. If,
instead, the workers return their own quantitative evaluations $\Ym$ of
the object quality, the evaluations carried out within the first round provide
useful information also for the second round.

In light of this discussion, it should not be surprising that the performance gain of algorithms exploiting 
workers' quantitative quality evaluations increases as the number of objects increases. 

Instead, when worker's bias come to play a significant role, it is a-priori much
less clear whether algorithms resorting on absolute (noisy) evaluations are
still preferable with respect to algorithms based on direct comparisons. This
happens because, while we may expect that the performance of the first tend to
degrade for effect of the worker's bias, the performance of the latter are
intrinsically insensitive to the worker's bias.

However, we wish to highlight that i) as already mentioned, the impact of worker
bias on the algorithm performance plays a marginal role, as all the contestant
objects are evaluated by the same worker, ii) the worker's bias can be rather
accurately estimated and subtracted when the worker has evaluated a sufficiently
large number of objects.  The combination of i) and ii) makes us pretty
confident that algorithms resorting on absolute quality estimations of objects
quality can be still competitive even when worker's bias is significant.
Results presented in the following in Section~\ref{sec:results} will confirm our
intuition.

\section{Answer Quantization}\label{sec:quanti}
Up to now, we have assumed that workers return unquantized (i.e.,
infinite-precision) noisy evaluations of object qualities.  This assumption is
unpractical in many scenarios, where instead workers' evaluations must belong to
a finite alphabet, i.e., they are quantized.  In this section, we discuss how
quantization can be effectively implemented in order to approach the performance
of the proposed unquantized algorithms.

From a system point-of-view, the key parameter of a quantizer is the cardinality
$L$ of the alphabet on which answers should be encoded, i.e., the number of
levels of the quantizer.  Given $L$, a specific quantization rule is
characterized by an $(L+1)$-dimensional vector of thresholds $\Zc = [z_1,\cdots,
  z_l, \cdots, z_{L+1}]$ with $z_1=-\infty<z_2 <z_3 < \cdots,
z_L<z_{L+1}=+\infty$ and an $L$-dimensional vector $\Wc = [w_1, \cdots, w_l
  \cdots, w_L]$ of representative values, satisfying $w_l \in (z_l, z_{l+1})$
If workers' answers are quantized,
  then the $j$-th answer to the evaluation of object $i$ can be modeled as
\[ a_{ij}^{(\rm q)} = \Qc(a_{ij})\]
 where $\Qc(x) = w_l$ whenever $ x \in (z_l, z_{l+1}]$.\footnote{Observe,
     however, that workers can be unaware of representative $\Wc$; every worker
     is just requested to express a satisfaction level in $\{1,\cdots, L\}$,
     which is obtained by comparing her own unquantized evaluation with
     thresholds $\Zc$. Therefore, our model perfectly matches the assumptions of
     \cite{GMhybrid}. } Notice that we consider here a \emph{fixed},
   non-adaptive quantizer, that is defined once and for all at the beginning,
   before any evaluation takes place.

In our context, the problem of optimal quantization can be formulated in terms
of the minimization of some distortion index between unquantized answers and
their quantized version.  The mean square error $\mathbb{E}[(a^{(\rm q)}-a)^2]$
represents a natural candidate for such distortion index, also because the
seminal work by Lloyd~\cite{lloyd} provides an efficient iterative algorithm for
the design of a quantizer that minimizes the mean square error.  Now, the
nontrivial question to be answered is: \emph{which are the answers whose
  distortion after quantization should be minimized?}  We list in the following
a few possible answers to such question.
\begin{itemize}
 \item Since we have $N$ objects whose quality values are i.i.d., each with pdf
   $f_q$, we may want to minimize the distortion on the answer relative to the
   generic object. With such a choice, the distribution with respect to which
   the mean square error is to be computed is $f_a^{(I)}=f_q * f_n$, where $*$
   is the convolution product and $f_n=\mathcal{N}(0,\sigma_b^2+\sigma^2)$ is the
   distribution of the workers' evaluation error.
 
 \item From a different perspective, since we are searching for the top-quality
   object, we should minimize the distortion on the answers associated to that
   object only, i.e., averaging with respect to $f_a^{(II)}=f_{q_{[1]}} * f_n$,
   $f_{q_{[1]}} $ being the a-priori distribution of the largest quality value.
 
 \item Taking an approach which is in-between the previous two, and considering
   that our target is to discriminate the best object from the others, we could
   aim at minimizing a weighted combination of the distortions relative to the
   ordered quality values.  This goal can be achieved by using as the answer
   distribution $f^{(III)}_a=\sum_{i=0}^{N-1} \alpha_i f_{q_{[i]}} * f_n$, where
   $f_{q_{[i]}}$ is the a-priori distribution for the $i$-th best object, and
   $\alpha_i$ is its associated weight, satisfying $\alpha_{i+1} \leq
   \alpha_{i}$.
 \end{itemize}

 We point out that, in order to better adapt to the
   distribution of objects to be evaluated, the quantizer should be
   redesigned at each round instead of just once at the
   beginning. This, however, entails a significant complexity increase of the
   algorithm. For the sake of simplicity in the following we will
   design the quantizer only once.
 
In the next section we show that the impact on algorithm performance of the
different possible choices for $f_a$ is pretty significant. Therefore, the
quantizer must be carefully designed. Finally, observe that $f^{(I)}_a$ and
$f^{(II)}_a$ can be obtained as particular cases of $f^{(III)}_a$ by setting
$\alpha_i=1/N$ for all $i$ in the first case, and $\alpha_1=1$, $\alpha_i=0$ for
$i>1$ in the second case.

\section{Results}
\label{sec:results}
In this section, we compare the performance of several algorithms that are
obtained by making different choices for the {\em fitness index}, the {\em
  allocation function}, the {\em termination rule}, and the \emph{quantizer}.
In Section~\ref{sec:results:simple} we first focus on a simple
  scenario where object qualities are equally spaced, the budget is unbounded
  (i.e., we can employ an unlimited number of evaluations), workers do not suffer
  from bias and provide unquantized answers. Then, in
  Sections~\ref{sec:results:bounded} and~\ref{sec:results:bias} we will show the
  effects of bounded budget and bias on system performance. We conclude by
  analyzing the effect of quantized answers in Section~\ref{sec:results:quantized}.

\subsection{Algorithm comparison in a simple scenario\label{sec:results:simple}}
In particular, we first focus on  algorithms with unquantized answers and
unbounded budget, and define:
\begin{itemize}
 \item the 'Greedy-Keep-Exact' (GKE) algorithm, employing the {\em exact max
   probability} as fitness index, {\em unbounded budget} as allocation function,
   and the {\em accuracy termination rule};
 
 \item the 'Greedy-Keep-Approximate' (GKA) algorithm, employing the {\em
   approximate max probability} as fitness index, the {\em unbounded budget} as
   allocation function, and the {\em accuracy termination rule};
 
 \item the 'Greedy-Remove-Approximate' (GRA) algorithm, employing the {\em
   approximate max probability with eliminations} as fitness index, the {\em
   unbounded budget} as allocation function, and {\em singleton contestant set}
   as termination rule.
\end{itemize} 

To reduce the space of parameters, we have always fixed $\pi_{\rm th,A}=\pi_{\rm
  th,E}\triangleq \pi_{\rm th}$. For the sake of comparison, the following
algorithms have been also considered.

\begin{itemize}
\item We have superimposed a classical tournament scheme to our previously
  described algorithms, obtaining a family of {\em Tournament}-$N_b$ (T-$N_b$)
  algorithms.  Specifically, in T-$N_b$ algorithms, the $N$ objects to be
  evaluated are first randomly partitioned into subgroups of size $N_b$ (for
  simplicity, we neglect rounding problems).  The GKA algorithm is then run to
  elect a winner for each of the object groups.  Only winners have access to the
  second stage, in which again objects are partitioned into subgroups of size
  $N_b$ and winners are elected for each subgroup. The process is iterated until
  only one winner is left.  We remark that our tournament schemes effectively
  exploit, at every stage, full information about the evaluations of competing
  objects collected at earlier stages.

\item A non-adaptive algorithm, which assigns to every object a fixed number of
  workers, referred to as Uniform algorithm (UA).

\item As a reference, we have also considered an unfeasible Genie-Aided (GA)
  algorithm, which has access to the identity of the two best competing objects,
  after a first initial round of evaluations, where every object receives one
  score.  Therefore, in the following rounds, the GA algorithm equally
  distributes workers only to the top two objects until the {\em accuracy
    termination rule} is met.  Observe that, by construction, the performance of
  the GA algorithm constitutes an upper bound for every feasible algorithm,
  since, as discussed in Section~\ref{sec:sco-vs-T}, it implements the optimal
  policy to find the best between two objects.

\end{itemize}

We start considering a simple scenario with $N=16$ objects whose qualities $x_i$
are equally spaced in the interval $[-1,1]$, so that the smallest difference
between quality values is $\Delta=\frac{2}{N-1}= \frac{2}{15}$.  The standard
deviation of the worker evaluation error has been set to $\sigma=\Delta/2$.

%%%%%%%%%%%%%%%%%%%%%%%%%%%%%
\begin{figure}[t]
\centering
\includegraphics[width=0.7\textwidth]{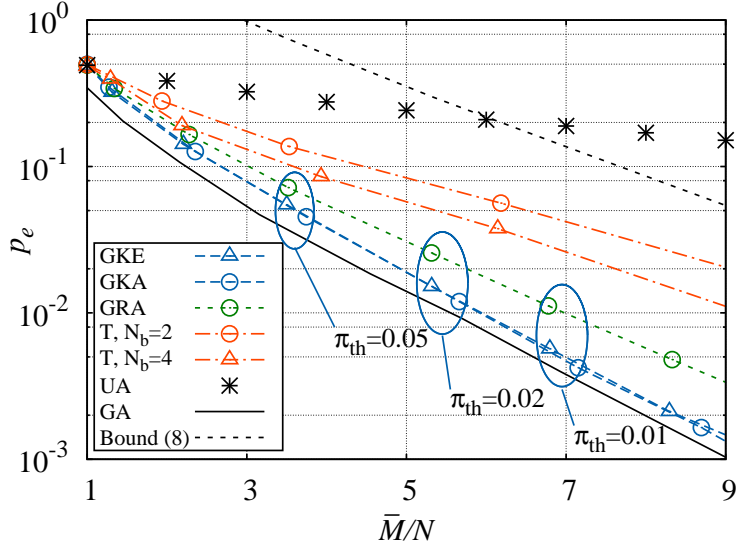}
\caption{Performance ($p_{\rm e}$ vs $\overline{M}/N$) of different algorithms
  with unquantized answers, no bias, and unbounded budget for equally spaced objects, 
  $N=16$, $\Delta/\sigma=2$.}
\label{fig1a}
\end{figure}

\begin{figure*}[t]
\centering
\includegraphics[width=0.7\textwidth]{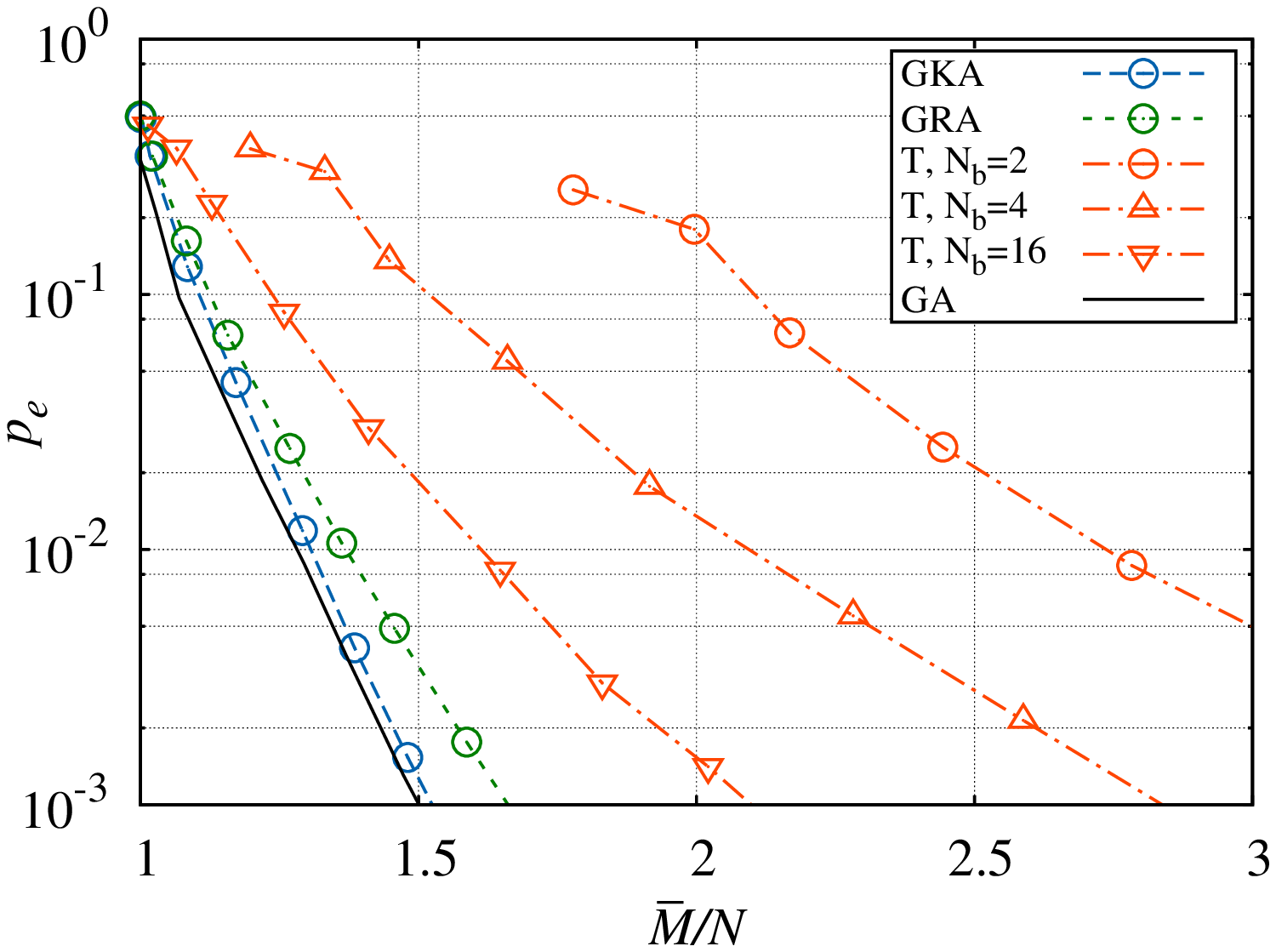}
\caption{Performance ($p_{\rm e}$ vs $\overline{M}/N$) of different algorithms
  with unquantized answers, no bias, and unbounded budget for equally spaced
  objects, $N=256$, $\Delta/\sigma=2$}
\label{fig1b}
\end{figure*}

\begin{figure*}[t]
\centering
\includegraphics[width=0.7\textwidth]{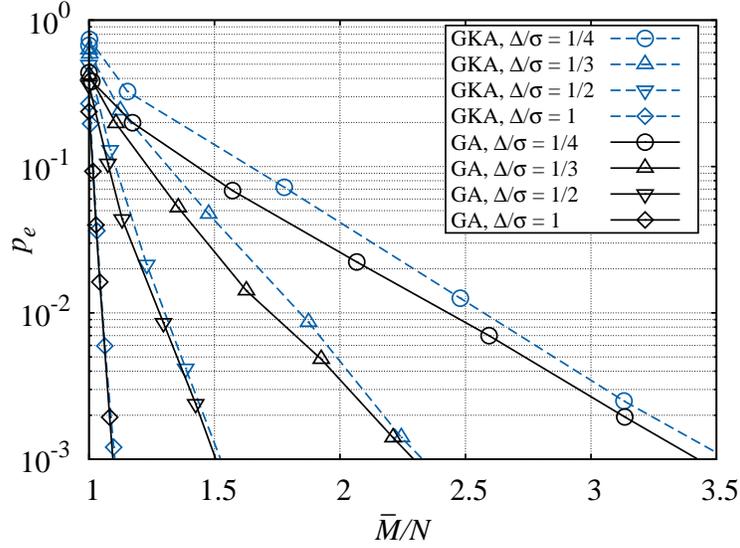}
\caption{Performance ($p_{\rm e}$ vs $\overline{M}/N$) of different algorithms
  with unquantized answers, no bias, and unbounded budget for equally spaced
  objects, $N=256$, and varying $\Delta/\sigma$.}
\label{fig1c}
\end{figure*}

Fig.~\ref{fig1a} shows the results obtained with the different proposed
algorithms, plotted in terms of error probability $p_{\rm e}$ versus the average
number of performed evaluations per object $\overline{M}/N$. We highlight that
the different trade-offs between $p_{\rm e}$ and $\overline{M}/N$ correspond to
different values of the threshold $\pi_{\rm th}$. Precisely, in
Fig.~\ref{fig1a}, blue circles group results obtained with the same values of $\pi_{\rm th}$.
Observe that the choice of $\pi_{\rm th}$ has a direct impact
on the expected error probability of the algorithm.  In particular, for the GKE
algorithm a simple analysis yields the following relationship between $\pi_{\rm
  th}$ and $p_{\rm e}$:
\beq p_{\rm e}=\sum_{i\neq \widehat{i}^*} \pi_i(\yv)\le
(N-1)\pi_{\rm th}
\eeq
More in general, for every algorithm, by decreasing
$\pi_{\rm th}$ we achieve a larger accuracy at the cost of employing more
resources.

From the results, the following observations can be made. 
\begin{itemize}
\item[i)]  Every adaptive algorithm performs much better than the uniform algorithm, employing on average the same amount of 
resources.

\item[ii)]  Greedy algorithms without  elimination perform  better than greedy algorithms with elimination. Observe, however that  
the latter are preferable in terms of  computational complexity,  since  in such schemes, at round $\ell$,  the fitness index 
 has to be computed only for objects in the contestant set $\Cc^{(\ell)}$, while in schemes without elimination it  has  to be computed for all objects.

\item[iii)] The selection of the approximate max probability as fitness index, in place of the exact max probability,   does not lead to any appreciable performance degradation, while having a significant  beneficial impact on the computational complexity of the algorithm
(this has been checked also for algorithms with elimination). 

\item[iv)] Tournament algorithms perform worse than our adaptive algorithms; furthermore, their performance tends to worsen as $N_b$ is reduced.

\item[v)] The performance of greedy algorithms without elimination is only marginally worse than that of the GA algorithm.
\end{itemize}

To gather more insight on the algorithm behavior, a further
performance comparison for the different schemes is reported in
Fig.~\ref{fig1b} for the case in which the number of objects is
increased to $N=256$ (object qualities $x_i$ are still equally
spaced in the interval $[-1,1]$, with $\sigma=\Delta/2$).  Observe
that the relative ranking among the algorithms does not change, but
the performance gap between algorithms tends to become more
significant.  In particular, tournament algorithms perform much worse
than GKA and GRA.  We remark that our results seem somehow
in contrast with findings in~\cite{GMhybrid}, where it has been shown
that tournament algorithms provide the best performance, for cases in
which users are only able to compare objects pairs.  To intuitively
grasp why they become inefficient in our context as $N$ increases,
notice that tournament algorithms waste a significant amount of
resources to discriminate among objects with similar quality
(accidentally placed in the same group), even when the quality of such
objects is much worse than top-quality values.  Observe that, also in
this case, GKA (whose performance is again practically
indistinguishable from GKE, not reported in Fig.~\ref{fig1b} for the
sake of readability) performs similarly to the GA algorithm.  This
proves the effectiveness of GKA in the considered scenarios.

To evaluate the impact of human evaluation errors on the overall
performance of algorithms, Fig.~\ref{fig1c} reports a performance
comparison between the GKA and GA algorithms for different values of
the parameter $\Delta/\sigma$.  Observe that $\Delta/\sigma$ plays an
important role: as the ratio $\Delta/\sigma$ decreases, more and more
resources are needed to meet the same error probability.  The
performance gap between the two algorithms is rather limited in all
cases. In particular, uniformly over all cases, the penalty cost in
terms of evaluations required to obtain the same error probability
does not exceed 10\%.  Once again, this confirms the effectiveness of
our approach for a broad range of scenarios where evaluation errors
have different impact.

\subsection{Effects of bounded budget\label{sec:results:bounded}}
Now, we move to scenarios in which the budget of allocations is
bounded, and we restrict our analysis to:
\begin{itemize}
 \item the 'bounded-Greedy-Keep-Approximate' (bGKA), employing the
   {\em approximate max probability} as fitness index, the {\em
     bounded budget} as allocation function, the {\em maximum budget
     achieved} or {\em accuracy termination rule};
 
\item the 'bounded-Greedy-Remove-Approximate', (bGRA) employing the
  {\em approximate max probability with eliminations} as fitness
  index, the {\em bounded budget} as allocation function, the {\em
    maximum budget achieved or singleton contestant list} as
  termination rule.  
\end{itemize}
As a reference, we report also the performance of the bounded version
of the GA algorithm (bGA), which again provides an obvious upper bound
to performance. Also in this case, to reduce the space of
  parameters, we fix $\pi_{\rm th}=\pi_{\rm th,A}=\pi_{\rm th,E}$. 

\begin{figure}[ht]
\centerline{\includegraphics[width=0.7\columnwidth]{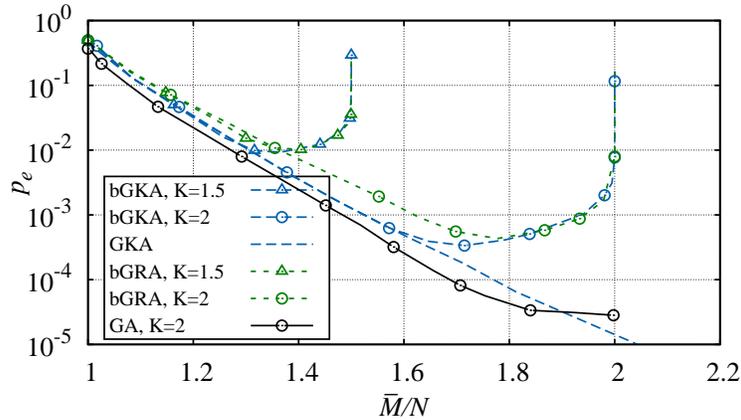}}
\vspace{-2ex}
\caption{$p_{\rm e}$ vs $\overline{M}/N$ for bGKA
  and bGRA, $N=256$ equally spaced objects.}\label{fig4}
\end{figure}

\begin{figure}[ht]
\centerline{\includegraphics[width=0.7\columnwidth]{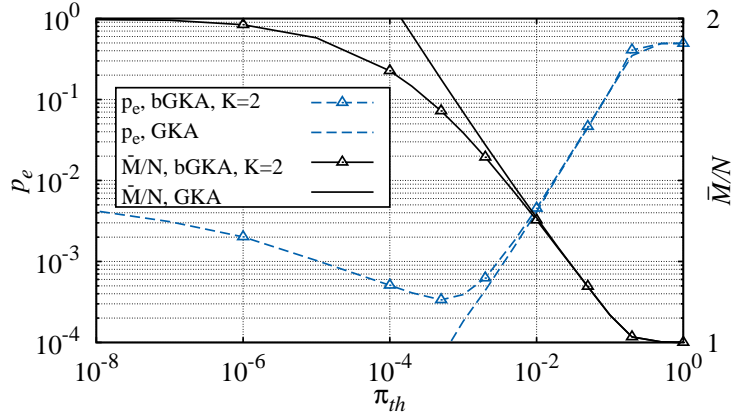}}
\vspace{-2ex}
\caption{$p_{\rm e}$ and $\overline{M}/N$ vs
  $\pi_{\rm th}$ for bGKA, $N=256$ equally spaced objects.}
\label{fig5}
\end{figure}

Fig.~\ref{fig4} compares the performance of different algorithms for
different values of the normalized budget $K=M_{\rm max}/N$. In
the same figure, we also report the performance of the unbounded
versions of the algorithms.  We observe that:
\begin{itemize}
\item[i)] the error probability for bGKA and bGRA now is not monotonic
  with respect to $\overline{M}/N$.  In particular, if the average
  number of evaluations $\overline{M}$ is sufficiently smaller than
  $M_{\rm max}$ (i.e., for sufficiently large values of $\pi_{\rm
    th}$), the performance of the bounded algorithms does not
  significantly differ from the respective unbounded version: this
  happens because the probability that the algorithm terminates for
  achieving maximum budget is negligible.  As we further reduce
  $\pi_{\rm th}$, increasing the required accuracy, the probability
  that the algorithm terminates because the maximum budget is achieved
  quickly increases, and the overall performance of the bounded
  algorithms degrades.  These effects can be better understood from
  Fig~\ref{fig5}, which reports the average number of evaluations per
  object, $\overline{M}/N$, and the error probability, as a function
  of the threshold $\pi_{\rm th}$, for the bGKA algorithm (similar
  considerations hold for bGRA).  Now, observe that $\overline{M}/N$
  monotonically increases as $\pi_{\rm th}$ decreases, since, by
  decreasing $\pi_{\rm th}$, the algorithm tends to be more
  conservative in excluding objects from receiving further
  evaluations. As a result, we distribute a larger number of
  allocations to objects with a quality quite distant from the
  maximum.  While the error probability behaves monotonically with
  respect to $\pi_{\rm th}$ (decreasing as $\pi_{\rm th}$ is
  decreased) in the case of an unbounded budget, in the case of
  bounded budget, choosing a value of $\pi_{\rm th}$ too small will
  lead to an inefficient distribution of the limited resources.

\item[ii)] As a consequence of i), only a limited range of $p_{\rm e}$
  values can be achieved in the bounded budget case.

\item[iii)]  Also in this case, bGKA  outperforms bGRA.
\end{itemize}

Now, we consider a scenario in which object qualities $x_i$ are
randomly drawn from a Gaussian distribution with zero mean and
standard deviation $\sigma_{\rm a}$.

\begin{figure}[ht]
\centerline{\includegraphics[width=0.7\columnwidth]{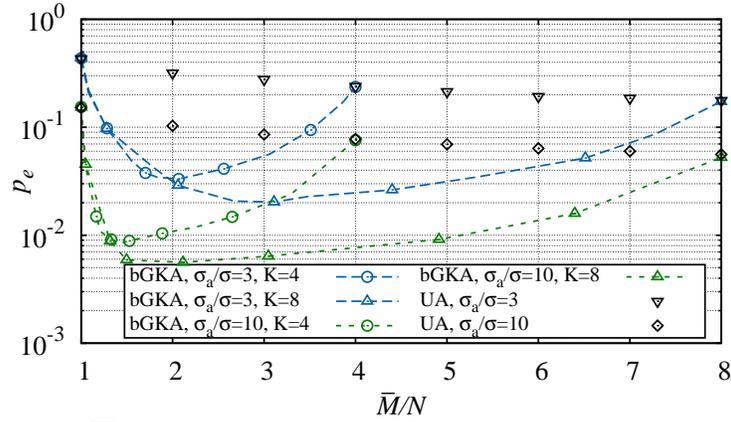}}
\vspace{-2ex}
\caption{$p_{\rm e}$ vs $\overline{M}/N$ for
  bGKA and Uniform, $N=256$. Object qualities are Gaussian.}
\label{fig6}
\end{figure}

\begin{figure}[ht]
\centerline{\includegraphics[width=0.7\columnwidth]{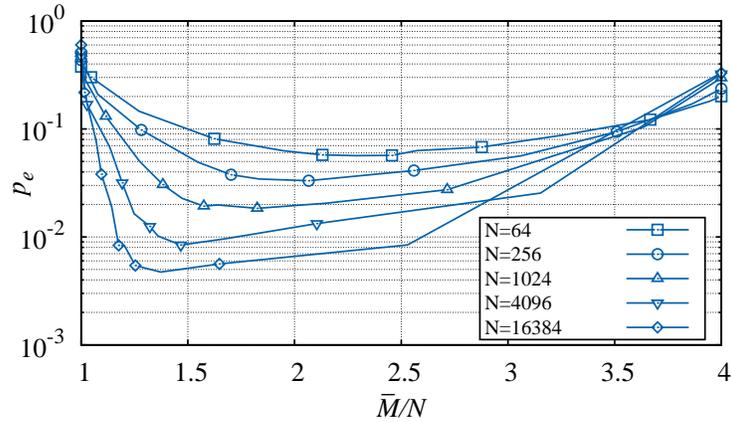}}
\vspace{-2ex}
\caption{$p_{\rm e}$ vs $\overline{M}/N$ as $N$ varies, for
  bGKA, $\sigma_{\rm a}/\sigma=3$ and $K=4$. Object qualities are Gaussian.}
\label{N_Gaussian_bounded}
\end{figure}

Fig. ~\ref{fig6} presents the performance of bGKA for the cases in
which $\sigma_{\rm a}/\sigma=3$ and
$\sigma_{\rm a}/\sigma=10$.  For the sake of comparison, the
figure also reports the performance of the uniform algorithm.

Fig.~\ref{N_Gaussian_bounded} shows the performance of
  bGKA algorithm for $\sigma_{\rm a}/\sigma=3$ and $K=4$ as the number
  of objects, $N$, varies. In particular:
  \begin{itemize}
    \item[i)] when the threshold   $\pi_{{\rm th},A}=0$ (i.e., when $\bar{M}/N=4$) the algorithm always
  assigns $K=4$ evaluations to each object.  Since as $N$
  increases, objects are denser, finding the maximum is more
  challenging and more prone to errors.
  \item[ii)] Instead, when the threshold
  significantly increases (left part of the figure), the algorithm is
  able to efficiently allocate the available evaluations on a small
  set of objects. Since the total budget increases with $N$, the number of evaluations 
assigned to the best objects increases with $N$ as well. 
Therefore, as $N$ grows the probability of exhausting the budget decreases. This has a
  beneficial effect on the overall error probability.
\end{itemize}

In conclusion we observe that: 
\begin{itemize}
\item[i)] when qualities are drawn from a
  Gaussian distribution, bGKA improves performance with respect to the
  uniform allocation algorithm that employs the same average budget of
  resources;

\item[ii)] from a qualitative point of view, bGKA exhibits a behavior
  which is pretty similar to the case where objects are equally spaced. By decreasing $\pi_{\rm th}$ we initially increase
  accuracy at the cost of increasing also the amount of employed
  resources.  However, beyond a given point, further decrements of
  $\pi_{\rm th}$ cause a loss of efficiency, worsening the overall
  performance of the algorithm;

\item[iii)] the performance of the algorithms heavily depends on the
  parameter $\sigma_{\rm a}/\sigma$, which can be regarded as a
  difficulty index for the problem.
\end{itemize}

\subsection{Effects of bias\label{sec:results:bias}}
\begin{figure}[ht]
\centerline{\includegraphics[width=0.7\columnwidth]{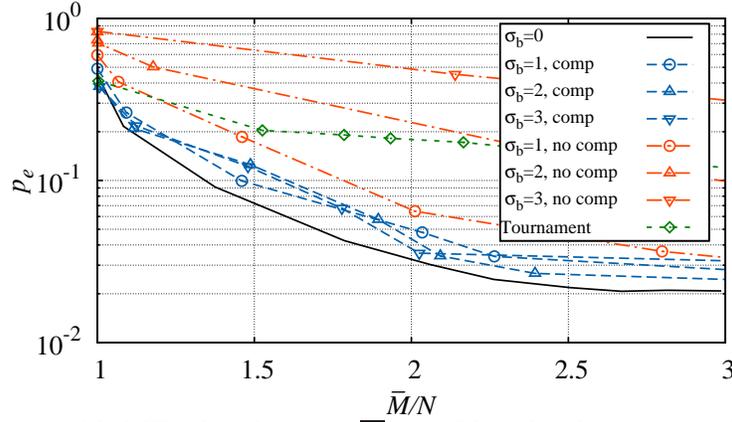}}
\vspace{-2ex}
\caption{Performance of the bGKA algorithm
  versus $\overline{M}/N$, with biased workers, $N=256$ objects, Gaussian-distributed qualities,
  $\sigma_{\rm a}=3$, $\sigma=1$, and for different values of the bias
  variance $\sigma^2_{\rm b}$.}
\label{fig:bias1}
\end{figure}

In Fig.~\ref{fig:bias1} we show the performance of the bGKA algorithm in the
case where workers suffer from bias. We test the algorithm for $N=256$ objects
with Gaussian-distributed qualities, $\sigma_{\rm a}=3$, $\sigma=1$, and for
different values of the bias variance $\sigma^2_{\rm b}$. Moreover, the budget
is bounded to $K=8$ and each worker can evaluate up to 256 objects.  Observe
that even if all the evaluations of a round can be performed by a single worker
in bGKA, by construction different objects are evaluated by different sets of
workers (indeed, only a subset of contestant objects is evaluated at every
round).  Therefore worker bias can potentially significantly affect the bGKA
performance if not properly compensated.

The solid line refers to the case $\sigma_{\rm b}=0$.
Instead, the dashed lines have been obtained for $\sigma_{\rm b}/\sigma=1,2,3$,
respectively. As expected, the bias leads to a moderate performance
degradation since it is effectively estimated and compensated.
We can also observe that the algorithm is weakly sensitive to the
bias variance. Indeed, the dashed lines are quite close to each other.

Instead the dash-dotted lines refer to the case where the algorithm does not
estimate and compensate for the bias. We see that in this scenario the
performance is significantly worsened.
For the sake of comparison we also
added the performance of the {\em T-2} algorithm for $\sigma_{\rm b}=0$
which shows considerable performance degradation with respect to the bGKA
algorithm with biased workers. Once again we remark that the {\em T-2}
algorithm is optimistic with respect to classical comparison algorithms as
explained in Section~\ref{sec:sco-vs-T}.
\begin{figure}[ht]
\centerline{\includegraphics[width=0.7\columnwidth]{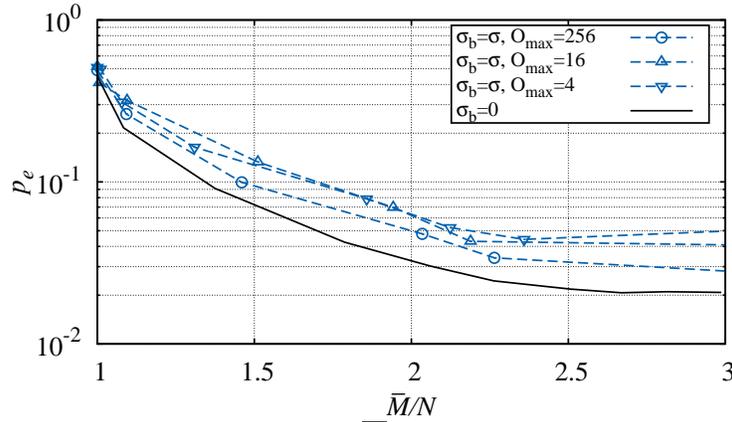}}
\vspace{-2ex}
\caption{Performance of the bGKA
  algorithm versus $\overline{M}/N$, with biased workers, $N=256$ objects,
  $\sigma_{\rm a}/\sigma=3$, $\sigma_{\rm b}/\sigma=1$, $K=8$, and for various
  values of $O_{\rm max}$.}
\label{fig:bias2}
\end{figure}
In Fig.~\ref{fig:bias2} we evaluate the
impact of the maximum number of objects that each worker can handle, $O_{\rm
  max}$ for $N=256$ objects with Gaussian-distributed qualities, $\sigma_{\rm
  a}/\sigma=3$, $\sigma_{\rm b}/\sigma=1$ and the budget is bounded to $K=8$. As
$O_{\rm max}$ decreases a larger number of workers are required to perform the
same task and, by consequence, the number of bias parameters to be estimated
also increases.  However we observe that the degradation is very limited even
when $O_{\rm max}=4$.

\subsection{Effects of quantization\label{sec:results:quantized}}
\begin{figure}[ht]
\centerline{\includegraphics[width=0.7\columnwidth]{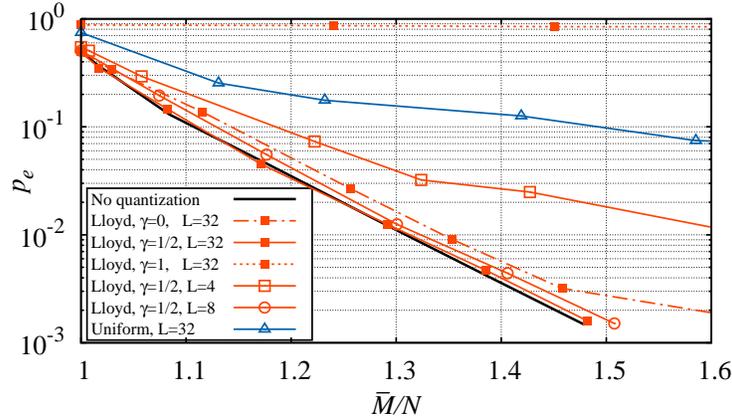}}
\vspace{-2ex}
\caption{Performance ($p_{\rm e}$ vs $\overline{M}/N$)
  of the GKA algorithm with unbounded budget and quantized workers' answers, for
  $N=256$ objects with equally spaced quality values. Workers' evaluations are
  affected by no bias.}
\label{fig:quantization_equallyspaced}
\end{figure}

In Fig.~\ref{fig:quantization_equallyspaced}, we test the impact of different
quantizers, in terms of $p_e$ versus $\overline{M}/N$, in the case where $N=256$
objects are equally spaced in the interval $[-1,1]$ and unbiased workers are
used. In the figure, the GKA algorithm with unbounded budget is employed for all
curves. Moreover, the standard deviation of the worker evaluation error has been
set to $\sigma = \Delta/2$. We recall that, for equally spaced objects,
$\Delta=2/(N-1)$ is the smallest distance between quality values.  The solid
line without markers represents the performance of the GKA algorithm without
quantization and is used as a benchmark. The line with triangle markers refers
to the performance obtained employing a quantizer with $L=32$ representative
values uniformly distributed in $[-1-2\sigma,1+2\sigma]$. We observe that,
despite of the high number of levels, uniform quantization significantly worsen
the error probability.  Instead, much better performance is achievable when the
quantizer design is more accurate. As an example, the solid line with filled
square markers refers to the case when $L=32$, and the quantizer is designed
according to the criteria in~\cite{lloyd}, over the answer distribution
$f^{(III)}_a=\sum_{i=0}^{N-1} \alpha_i f_{q_{[i]}} * f_n$, where
$\alpha_i=\gamma^i$, and $\gamma=1/2$. This quantizer, labeled ``Lloyd'' in the
legend, provides performance close to the unquantized case. In general, we
observe that the performance is quite insensitive to the design parameter
$\gamma$ except if its value is close to the extreme point
$\gamma=1$. Fig.~\ref{fig:quantization_equallyspaced} also shows the performance
of Lloyd's quantizers with $\gamma=1/2$ and number of levels $L=4,8$.  We
observe that an accurately designed quantizer with only $L=8$ levels is enough
to provide performance close to the unquantized case.
  
\begin{figure}[ht]
\centerline{\includegraphics[width=0.7\columnwidth]{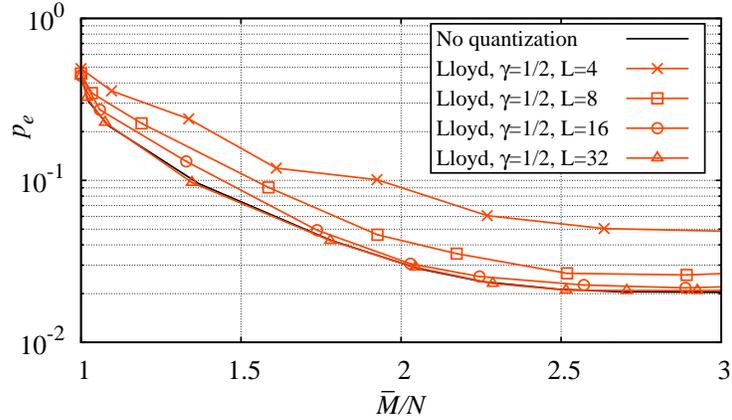}}
\vspace{-2ex}
\caption{Performance of the GKA algorithm
  with bounded budget, quantized workers' answers, for $N=256$ objects with
  Gaussian-distributed qualities.  Workers' evaluations are affected by no
  bias.}
\label{fig:quantization_gaussian}
\end{figure}

Fig.~\ref{fig:quantization_gaussian} refers to the case of $N=256$ objects
with Gaussian-distributed qualities, the budget is bounded to $K=3$, and workers'
answers are quantized. The quantizer is designed according to the Lloyd's
algorithm over the weighted distribution $f^{(III)}_a=\sum_{i=0}^{N-1}
\alpha_i f_{q_{[i]}} * f_n$, with $\alpha_i=\gamma^i$, and $\gamma=1/2$.  In the
figure, the solid line refers to the unquantized case, while the lines with
markers refer to the case where quantization is employed with $L=4,8,16,32$
levels. Also in this scenario, we observe that $L=8$ quantization levels are
enough to provide performance close to the unquantized case.

\begin{figure}[ht]
\centerline{\includegraphics[width=0.7\columnwidth]{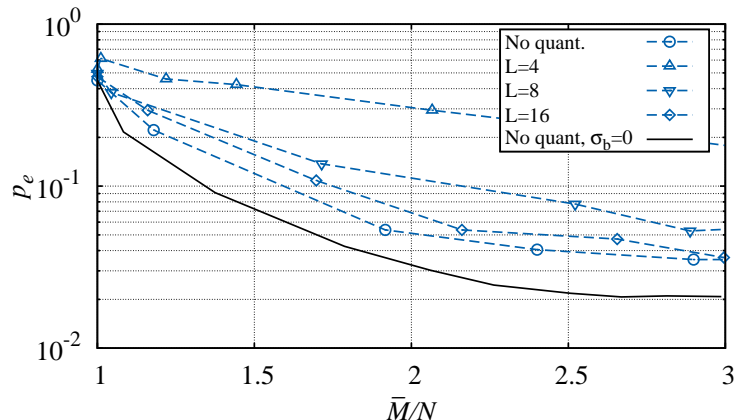}}
\vspace{-2ex}
\caption{Performance of the
  GKA algorithm with bounded budget, quantized workers' answers, for $N=256$
  objects with Gaussian-distributed qualities and $O_{\rm max}=256$. Workers'
  evaluations are affected by bias.}
\label{fig:bias256}
\end{figure}

\begin{figure}[ht]
\centerline{\includegraphics[width=0.7\columnwidth]{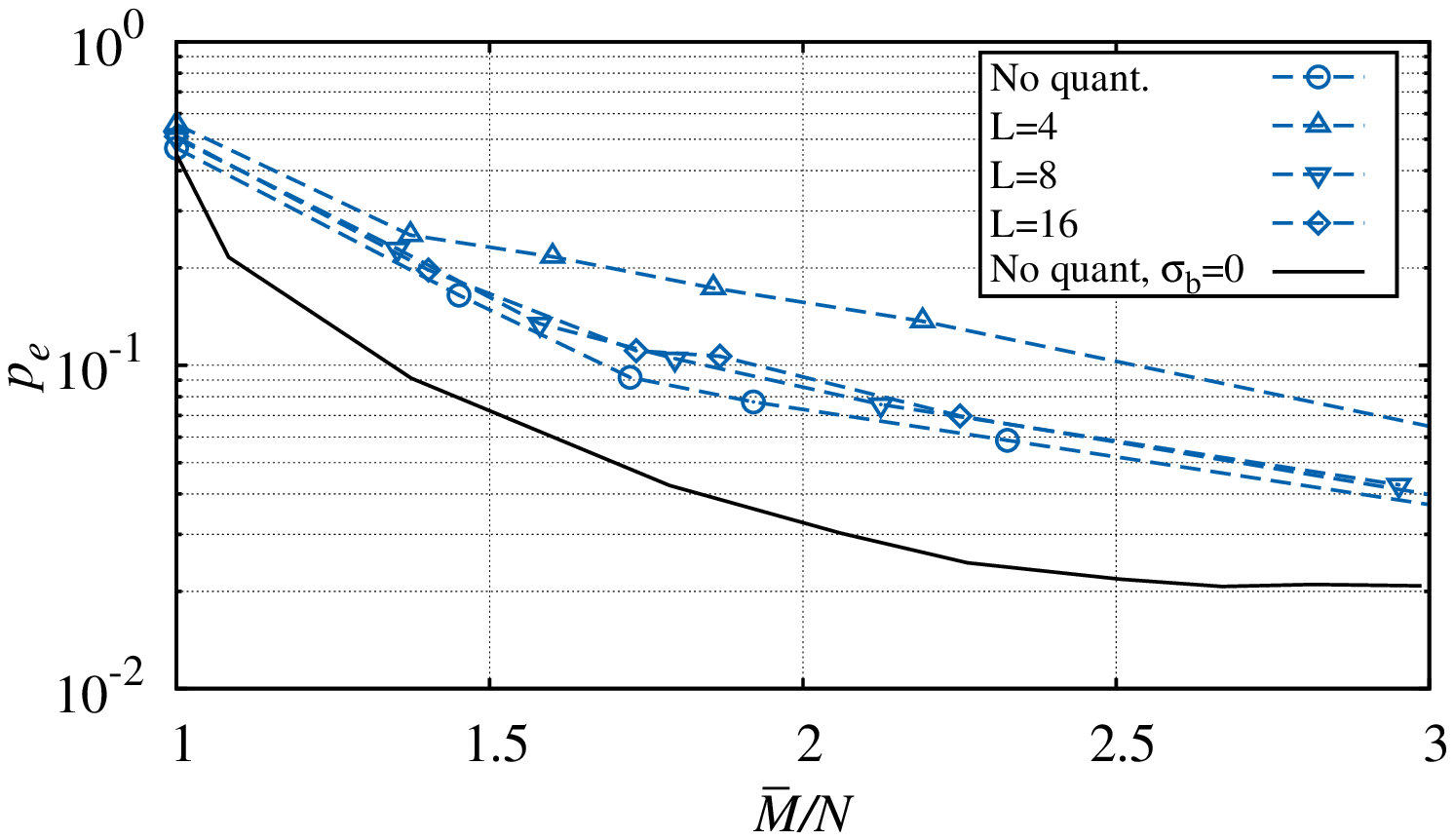}}
\vspace{-2ex}
\caption{Performance
  of the GKA algorithm with bounded budget, quantized workers' answers, for
  $N=256$ objects with Gaussian-distributed qualities and $O_{\rm
    max}=16$. Workers' evaluations are affected by bias.}
\label{fig:bias16}
\end{figure}

In Figs.~\ref{fig:bias256} and~\ref{fig:bias16} we combine the effects
of bias, quantization and bounded budget with $N=256$ objects with
Gaussian-distributed qualities, $\sigma_{\rm a}/\sigma=3$,  $\sigma_{\rm b}/\sigma=1$ and $K=8$.
The figures show the error probability $p_e$ versus the $\bar{M}/N$ for
$L=4,8,16$ quantization levels. The cases with unquantized answers and unbiased
workers are also reported.

Specifically in Fig.~\ref{fig:bias256} we set $O_{\rm max}=256$ so that at each
round a single additional worker (and hence a single bias parameter to
be estimated) is required, while in Fig.~\ref{fig:bias16} we considered the more
challenging scenario where $O_{\rm max}=16$. In the latter case, at each round, up
to 16 workers are required.

We observe that in both cases 16 quantization levels are enough to limit the
performance degradation. Surprisingly, for $L=4$ performance worsens with
increasing $O_{\rm max}$. This is due to complex interaction between the bias
estimation algorithm and the quantizer.

\begin{figure}[ht]
\centerline{\includegraphics[width=0.7\columnwidth]{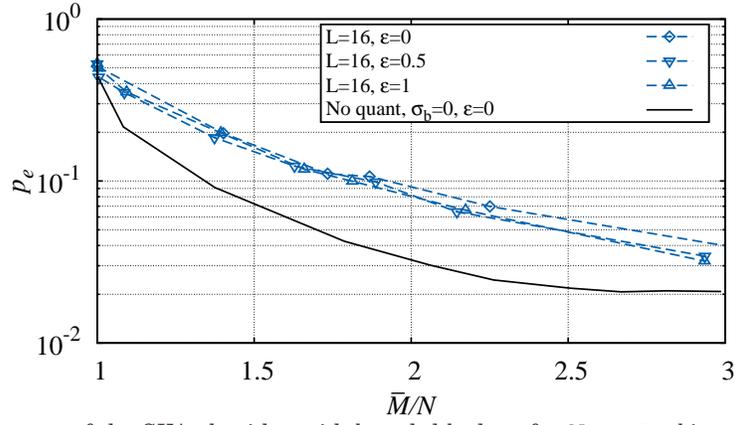}}
\vspace{-2ex}
\caption{Performance
  of the GKA algorithm with bounded budget, for $N=256$ objects with
  Gaussian-distributed qualities $L=16$ and $O_{\rm max}=16$. Workers'
  evaluations are affected by bias and their evaluation variance is
  uniformly distributed with support $[(1-\epsilon)\sigma^2,(1+\epsilon)\sigma^2]$.}
\label{fig:bias16_variances}
\end{figure}

Finally in Figure~\ref{fig:bias16_variances} we
  investigate the effect of workers with different skills on the
  system performance. Precisely each worker is characterized by a
  random evaluation variance, uniformly distributed in
  $[(1-\epsilon)\sigma^2,(1+\epsilon)\sigma^2]$ where
  $0\le \epsilon \le 1$. The system parameters are the same as in
  Figure~\ref{fig:bias16} and we consider $L=16$ quantization
  levels.  It can be observed that the impact of random variances is
  very limited. As a matter of fact, an algorithm able to estimate and
  exploit such variances would provide better performance at a price of
  a complexity increase.  Such an algorithm could be based on belief
  propagation as in~\cite{Oh}.  However, the results depicted in the
  figure show that our algorithm is robust and able to handle workers
  with different skills.

\section{Design Considerations}
\label{sec:DC}

In this section we address two important issues: we evaluate the
computational complexity of our algorithms, and we discuss how to set
algorithm parameters (and in particular $\pi_{\rm th}$). For the sake
of brevity, we restrict our investigation to bGKA, which turns out to
be the best-performing algorithm.

For what concerns computational complexity, at round $\ell$ and in
  presence of bias in workers' answers the bGKA algorithm must
\begin{enumerate}
\item[i)] compute
  $\muv_{\qv,\betav}^{(\ell)}$ (this operation has complexity $O(N^3)$ since it
  requires the computation of the inverse of a $N\times N$ matrix);
\item[ii)] compute a fitness index $\phi^{(\ell)}_i$ for every object (this
  requires $O(N)$ operations since the {\em approximate max probability} can be
  computed by exploiting \eqref{approx-fit});
\item[iii)] compare $\phi^{(\ell)}_i$ with $\pi_{\rm th}$, for $i=1,\ldots,N$,
  in order to decide whether to allocate extra workers (again this
  requires $O(N)$ operations).
\end{enumerate}
In the final round, if the number of objects that pass the threshold exceeds the
residual budget, the execution of an extra task is necessary: objects must be
sorted in order of their fitness index (the cost of sorting is notoriously
$O(N\log N))$.  Observing that $M_{\rm max}$ is an obvious upper bound to the
number of rounds, it turns out that the overall complexity of bGKA can be
upper-bounded by $O(M_{\rm max}N^3)$.

Instead, when bias is not estimated, matrix inversion is not
required. In such scenario, the complexity of step i) becomes $O(N)$, thus the
overall complexity is $O(M_{\rm max}N+N\log N)=O(M_{\rm max}N)$ in consideration
of the fact that by construction $M_{\rm max}\ge N$.

As regards the setting of the algorithm parameters, which in the case of bGKA
are $\pi_{\rm th}$, $O_{\rm max}$ and $M_{\rm max}$, we observe that the choice
of the value for $M_{\rm max}$ normally depends on economical or
application-oriented considerations, whose discussion is beyond the scope of
this paper.  Given the value of $M_{\rm max}$, it is possible to tune the
algorithm by selecting the value for $\pi_{\rm th}$, which should be set in
order to minimize the error probability.  For a careful setting of $\pi_{\rm th}$,
a preliminary parametric analysis of the algorithm performance is
necessary to estimate the key system parameters (such as the pdf $f_q$ of object
qualities, the bias variance, $\sigma^2_b$, and the variance $\sigma$ of the
workers' quality estimation errors).

\section{Possible extensions to the top-$k$ object selection}
Although we have described our algorithms in a particular setting, our approach can be 
easily extended to more general scenarios. In particular, it is possible to generalize 
the approach proposed in this paper to the problem of
finding the $k$ top-quality elements within a large collection of objects through 
crowdsourcing algorithms. Indeed, as an extension of $\pi_i$, we can define 
$\pi_i^{(k)}$ as the probability for the $i$-th object to be among the $k$ best. 
Then, a possible algorithm could divide objects into three categories: 
i) those that are with high probability within the $k$ best, 
ii) those that with high probability are not within the $k$ best, and 
iii) the remaining ones. 
This can be done by defining \emph{two} thresholds instead of only one. 
The mechanism of allocation of new evaluations to objects is a straightforward extensions 
of the one described in the paper. 
More precisely, new evaluations are requested at each round only for objects of the third category.

\section{Concluding Remarks}
\label{sec:conclusions}
In this paper, we have studied the problem of finding the top-quality element
within a large collection of objects, resorting to human evaluations affected by
noise and by bias. Differently from previous works, our study started assuming
that unquantized scores are returned by the evaluators, and highlights the
potential advantages of such approach. We have shown that bias can be estimated
and compensated with an affordable additional complexity.

Then we have shown how to properly design
quantized schemes whose performance is very close to their ideal unquantized
counterparts, provided that a reasonable number of quantization levels is
assigned to workers' answers.

\bibliographystyle{ACM-Reference-Format-Journals}
%\bibliographystyle{abbrv}
%\begin{bibliography}

\end{document}